\documentstyle[prb,preprint,aps]{revtex}

\begin{document}
\draft
\title{Transport in quantum wells in the presence of interface roughness}
\author{Chung-Yu Mou\cite{i} and Tzay-ming Hong\cite{ii}}
\address{Department of Physics, National Tsing-Hua University, Hsinchu, Taiwan 30043,%
\\
Republic of China}
\date{\today }
\maketitle

\begin{abstract}
The effective Hamiltonian for two dimensional quantum wells with rough
interfaces is formally derived. Two new terms are generated. The first term
is identified to the local energy level fluctuations, which was introduced
phenomenologically in the literature for interface roughness scattering but
is now shown to be valid only for an infinite potential well or Hamiltonians
with one single length scale. The other term is shown to modulate the
wavefunction and cause fluctuations in the charge density. This will further
reduce the electron mobility to the magnitude that is close to the
experimental result.
\end{abstract}


\section{Introduction}

The vast interest in the physics of charge transport in two-dimensional
quantum wells stems from potential applications in new devices and
subsequent integration with Si-based chip technology. Experimentally, it is
known that the charge transport inside a quantum well is strongly affected
by the quality of the well. In particular, it is believed that the interface
roughness is inherent to the quantum well systems and plays an important
role for wells at low temperature with small well widths\cite{sakaki}. On
the theoretical side, starting from the seminar work by Kardar, Parisi, and
Zhang\cite{KPZ}, a large effort has been devoted to understand the
morphology of thin film growth\cite{review} during the past decade.
Nevertheless, these works only characterize long wavelength properties of
the surface roughness, there is no systematic attempt to investigate how the
electronic properties, such as the charge transport, are affected by the
surface roughness.

The study of the effects of surface roughness on the electronic transport
properties has a long history, tracing back to the work by Prange and Nee 
\cite{prange} on magnetic surface states in metals. Later, a more complete
model was reconsidered by Ando\cite{Ando1}. Quite often, these works are
summarized phenomenologically by introducing a local energy-level
fluctuation term in potential: $(\partial E/\partial L)\Delta ({\bf r})$,
where $E$ is the energy eigenvalue of the electron, $L$ is the averaged well
width and $\Delta ({\bf r})$ is the local change of quantum well width. Such
phenomenology finds its natural application in interpreting the
photoluminescence data of GaAs/AlAs quantum wells\cite{sakaki,tanaka1}. In
this case, it has been established that for temperature less than 80 K, the
linewidth of photoluminescence is mainly determined by the local energy
level fluctuations. Transport properties of two-dimensional (2D) carrier
gases at Si/SiO$_{2}$ interfaces and in semiconductor quantum wells are also
shown to be strongly affected by the interface roughness\cite
{sakaki1,Laikhtman,Penner}. In these studies, theoretical mobility is also
calculated based on the assumption that the local energy level fluctuation
is the dominant effect. It is known, however, that the experimentally
observed mobility can not be explained solely by the roughness roughness. In
some parameter regime, one has to introduce, for example, phenomenologically
negative impurity charge to account for the extra reduction of the mobility
observed in experiments\cite{Penner,Emeleus}.

In this work, we shall systematically investigate the effects of surface
roughness. Our starting point is an averaged version of the Hamiltonian
specialized to the quantum well configuration. In an expansion in $\Delta (%
{\bf r})/L$, two lowest order terms are considered . The first term is
identified to the local energy level fluctuations $(\partial E/\partial
L)\Delta ({\bf r})$. This term represents the mismatch effect of the energy
band. It was introduced phenomenologically in the literature for interface
roughness scattering, but is now shown to be valid only for an infinite
potential well or Hamiltonians with one single length scale. The other term
is shown to modulate the wavefunction and cause fluctuations in the charge
density. This will further reduce the electron mobility.

The rest of the paper is organized as follows: In Section 2, we lay down the
formulation of the effective Hamiltonian for two dimensional quantum wells.
Some general features of the effective Hamiltonian are discussed. In
particular, we expand the Hamiltonian to the order the $\Delta ({\bf r})/L$
and discuss its effects on the basis of single particle states. In Section
3, we consider the fluctuation of the charge density caused by the
wavefunction modulation, and study its consequence on the electron mobility
to the order of $\Delta ({\bf r})/L$. In the final section, we examine the
validity of our approach and conclude.

\section{Theoretical formulation}

Let us consider a generic quantum well specified by two interfaces at $%
z=z_{+}({\bf r})$ and $z=z_{-}({\bf r})$, where ${\bf r}$ $=(x,y)$ is a two
dimensional vector. The average distance between the two surfaces is $L$
(see Fig.1). For simplicity, we shall impose the hard wall condition on the
interfaces. Our formulation is easily generalized to the case when the
potential well is finite. To investigate the effects that are due to the
interface roughness, it is convenient to do a transformation that maps $%
z_{+}({\bf r})$ to $L$ and $z_{-}({\bf r})$ to $0$. This transformation is
easily implemented by 
\begin{equation}
z^{\prime }=\frac{Lz}{\Delta ({\bf r})+L}-\frac{L\,z_{-}({\bf r})}{\Delta (%
{\bf r})+L},\;{\bf r}^{\prime }={\bf r,}
\end{equation}
where $\Delta ({\bf r})\equiv $ $z_{+}({\bf r})-$ $z_{-}({\bf r})-L.$ After
transformation, the wavefunction can be generically expressed by $\Psi
_{n}=\psi _{n}(x,y)\sin \left( \frac{n\pi }{L}z\,^{\prime }\right) /\sqrt{%
\frac{L+\Delta ({\bf r})}{2}}$which satisfies the normalization condition: $%
\int_{0}^{L+\Delta }dz\int d{\bf r\,|}\Psi _{n}|^{2}=1$.

For typical quantum wells, the Fermi wavelength is about 400${\rm \AA }$. If
L is less than 340${\rm \AA }$, there will be no band crossing at low
temperatures, and we can take the average along z direction with respect to
a given subband, i.e., average with respect to $\sin \frac{n\pi }Lz^{\prime
} $ ($n$ will be taken to be one). In other words, the more appropriate
Hamiltonian to work with is defined by 
\begin{equation}
H_n=\langle {\rm \hat{H}}\rangle \equiv \frac{\int_0^Ldz^{\prime }\sin \frac{%
n\pi }Lz\,^{\prime }{\rm \hat{H}}\sin \frac{n\pi }Lz^{\prime }}{%
\int_0^Ldz^{\prime }\sin ^2\frac{n\pi }Lz^{\prime }\,}.  \label{first}
\end{equation}
After the averaging, we find that an extra potential $\delta V$ is
introduced to $H_n=\frac{-\hbar ^2}{2m}\nabla ^2+\delta V$ where the
Laplacian is over the $(x,y)$ directions and 
\begin{eqnarray}
-\frac{2m}{\hbar ^2}\delta V &=&\alpha _n\frac{(\nabla A)^2}{A^2}+\beta _n%
\frac 2A\nabla A\cdot (\nabla B-\frac BA\nabla A)  \nonumber \\
&&+\gamma _n\left[ A^2+\left( \nabla B-\frac BA\nabla A\right) ^2\right]
+\delta _n\left( \frac 2A\nabla A\cdot \nabla +\frac 1A\nabla ^2A\right) .
\label{potent}
\end{eqnarray}
Here $A\equiv \frac L{\Delta ({\bf r})+L}$, $B\equiv -\frac{L\,z_{-}({\bf r})%
}{\Delta ({\bf r})+L}$, $\alpha _n\equiv \langle z^{\prime 2}\frac{\partial
^2}{\partial z^{\prime 2}}\rangle _n$, $\beta _n\equiv \langle z^{\prime }%
\frac{\partial ^2}{\partial z^{\prime 2}}\rangle _n$, $\gamma _n\equiv
\langle \frac{\partial ^2}{\partial z^{\prime 2}}\rangle _n$ and $\delta
_n\equiv \langle z^{\prime }\frac \partial {\partial z^{\prime }}\rangle _n$%
. It is easy to show that $\delta _n=-1/2$ is generally true.

If we expand $\delta V$ to the linear order of $\Delta /L$ and keep only up
to $O(z_{-})$, we obtain 
\begin{equation}
\delta V=E_{n}-\frac{2E_{n}}{L}\Delta ({\bf r})+\delta _{n}\frac{\hbar ^{2}}{%
m}\left( \frac{1}{L}{\bf \nabla }\Delta ({\bf r})\cdot {\bf \nabla +}\frac{1%
}{2L}\nabla ^{2}\Delta ({\bf r})\right) -\beta _{n}\frac{\hbar ^{2}\nabla
(z_{+}-L)\cdot \nabla z_{-}}{mL},  \label{potential1}
\end{equation}
where we have identified $-\gamma _{n}\hbar ^{2}/2m$ as $E_{n}$. Note that
the resulting Hamiltonian is invariant under reflection: $L-z_{-}\rightarrow
z_{+}$ and $L-z_{+}\rightarrow z_{-}$. Let us first put the last term in the
right place. For this purpose, we consider the two point correlation
function of $\Delta ({\bf r})$%
\begin{eqnarray*}
\langle \Delta ({\bf r})\Delta ({\bf r}^{\prime })\rangle  &=&\langle (z_{+}(%
{\bf r})-L)(z_{+}({\bf r}^{\prime })-L)\rangle +\langle z_{-}({\bf r})z_{-}(%
{\bf r}^{\prime })\rangle  \\
&&-\langle (z_{+}({\bf r})-L)z_{-}({\bf r}^{\prime })\rangle {\bf -\langle (}%
z_{+}({\bf r}^{\prime })-L{\bf )}z_{-}({\bf r})\rangle .
\end{eqnarray*}
If the $z_{+}({\bf r})$ and $z_{-}({\bf r})$ are uncorrelated, the cross
terms in the above equation vanish. In this case, if one assumes that the
surfaces described by $z_{+}({\bf r})$ and $z_{-}({\bf r})$ are
statistically the same, one obtains that $\langle (z_{+}({\bf r})-L)(z_{+}(%
{\bf r}^{\prime })-L)\rangle =\langle z_{-}({\bf r})z_{-}({\bf r}^{\prime
})\rangle =\langle \Delta ({\bf r})\Delta ({\bf r}^{\prime })\rangle /2$. In
other words, both $z_{+}({\bf r})-L$ and $z_{-}({\bf r})$ are of the order
of $\Delta ({\bf r})$. Therefore, the last term in Eq. (\ref{potential1}) is
of higher order and can be neglected. In addition to the special case when $%
z_{+}(r)-L$\ and $z_{-}(r)$\ are correlated, for instance, if either
interface is smooth but tilted, i.e., $\nabla z$\ is finite, this $\nabla
(z_{+}-L)\cdot \nabla z_{-}$\ term will also need to be considered.

In general, the second term in Eq.(\ref{potential1}) has no definite
relation with $E_{n}$. When $L$ is the only length scale in the Hamiltonian
(e.g., an infinite potential well), $E_{n}$ has to be proportional to $%
1/L^{2}$ and $-\frac{2E_{n}}{L}=\frac{\partial E_{n}}{\partial L}$. Eq.(\ref
{potential1}) can then be written as 
\begin{equation}
\delta V=E_{n}+\frac{\partial E_{n}}{\partial L}\Delta ({\bf r})-\frac{\hbar
^{2}}{2m}\left[ \frac{1}{L}{\bf \nabla }\Delta ({\bf r})\cdot {\bf \nabla +}%
\frac{1}{2L}\nabla ^{2}\Delta ({\bf r})\right] .  \label{potential}
\end{equation}
Physically it becomes clear that the second term in Eq.(\ref{potential})
describes the local energy-level fluctuation, which was introduced
phenomenologically in the literature for interface roughness scattering\cite
{Laikhtman,Penner} but is now formally derived and shown to be valid only
for an infinite potential well or any other potential with only one length
scale. It is easy to check that, to the first order in ordinary perturbation
theory, the third term in Eq.(\ref{potential}) does not contribute to the
scattering matrix for single-particle states. However, as we shall derive in
below, a closer investigation shows that this is not correct. Indeed, when
this term is combined with the kinetic energy, it becomes 
\begin{equation}
H^{\prime }=-\frac{\hbar ^{2}}{2m}\left( {\bf \nabla +}\frac{\nabla \Delta (%
{\bf r})}{2L}\right) ^{2}.
\end{equation}
(a second order term has been neglected). Obviously, its effect is to
modulate the wavefunction for every particle by 
\begin{equation}
\Psi ({\bf r})\rightarrow \Psi ({\bf r})\exp (-\frac{\Delta ({\bf r})}{2L}).
\label{exp}
\end{equation}
Note that this result is {\it independent} of the depth of the well. There
are two consequences: First, it causes fluctuations in the charge density.
This is a many-particle effect and will suppress the mobility. We will
analyze it in the next section. Secondly, even for single-particle state, it
implies that the transmission probability is not one when each particle
passes by a step in the interface. When combined with the Landauer formula
\cite{Landauer}, it indicates that the mobility will be further reduced. We
will analyze this effect in the final section.

In summary, to the first order in $\Delta ({\bf r})$, the surface roughness
introduces local energy level fluctuations and local wavefunction modulation
for single particle states. Since the surface roughness modulates the single
particle wavefunction in a coherent way, the wavefunction modulation does
not introduce new scattering. In the single particle level, the effect of
surface roughness is entirely contained in the local energy level
fluctuations. We shall see in the next section that when we include
many-particle interactions, the wavefunction modulation becomes as important
as the energy level fluctuations.

\section{Many-particle effect}

In this section, we discuss the many-particle effect that is due to the
wavefunction modulation. We shall demonstrate its effect on the calculation
of electron mobility. The change of the wavefunction induces a local
modulation in the density of electrons: 
\[
n({\bf r},z^{\prime })=n^{\prime }\left( {\bf r}\right) \frac{\sin ^2(\frac{%
n\pi }Lz^{\prime })}{(L+\Delta ({\bf r}))/2}\exp (-\frac{\Delta ({\bf r})}L) 
\]
with the understanding that the normalization is done with respect to $z$.
Here $n^{\prime }\left( {\bf r}\right) $ is the 2-D electron density after
being perturbed by the local energy level fluctuations, $\frac{\partial E_n}{%
\partial L}\Delta \left( {\bf r}\right) $. It is easy to show that

\[
n^{\prime }\left( {\bf r}\right) =n_{0}\left\{ 1+\frac{2m}{\pi \hbar
^{2}k_{F}^{2}}\frac{\partial E_{n}}{\partial L}\int\limits_{k\leq k_{F}}d^{2}%
{\bf k}\sum_{{\bf p}}\frac{\left[ \Delta ({\bf k}-{\bf p})\cdot e^{i({\bf p}-%
{\bf k})\cdot {\bf r}}+h.c.\right] }{k^{2}-p^{2}}\right\} 
\]
where $n_{0}=k_{F}^{2}/2\pi $ is the equilibrium electron density at two
dimensions and $h.c.$ denotes a Hermitian conjugate of the previous term. We
shall assume that the density of positive charge background remains
unchanged so that the local charge modulation is entirely due to electrons.
The change of charge density is 
\begin{equation}
\delta \rho ({\bf r},z)=en({\bf r},z^{\prime })\theta (z-z_{-})\theta
(L+\Delta -z)-en_{0}\frac{\sin ^{2}\left( \frac{n\pi }{L}z\right) }{L/2}%
\theta \left( z\right) \theta \left( L-z\right) .
\end{equation}
For convenience, we shall assume $z_{-}=0$ and neglect the curvature effect
due to the roughness (for instance, the special case when both interfaces
fluctuate while their spacing remains $L$). To the first order in $\Delta (%
{\bf r})$, the total electric potential $\delta \phi $ satisfies 
\begin{eqnarray}
-(\nabla ^{2}+\frac{\partial ^{2}}{\partial z^{2}})\delta \phi ({\bf r},z)
&=&4\pi \rho _{ind}+4\pi en_{0}\frac{\sin ^{2}(\frac{n\pi }{L}z)}{L/2}%
\left\{ -2\left[ 1+\frac{n\pi z}{L}\cot \left( \frac{n\pi }{L}z\right) %
\right] \frac{\Delta ({\bf r})}{L}\right.   \nonumber \\
&&\left. +\frac{m}{\pi ^{2}\hbar ^{2}n_{0}}\frac{\partial E_{n}}{\partial L}%
\int\limits_{k\leq k_{F}}d^{2}{\bf k}\sum_{{\bf p}}\frac{\left[ \Delta ({\bf %
k}-{\bf p})\cdot e^{i({\bf p}-{\bf k})\cdot {\bf r}}+h.c.\right] }{%
k^{2}-p^{2}}\right\} \theta (z)\theta (L-z),  \label{ming}
\end{eqnarray}
where $\rho _{ind}$ is the induced charge density. The associated scattering
matrix within a given subband is given by 
\begin{eqnarray}
\delta M({\bf q}) &=&\langle {\bf k}|\delta V|{\bf q-k}\rangle _{n} 
\nonumber \\
&=&\int d^{2}{\bf r\;}e^{i{\bf q\cdot r}}\;\frac{\int_{0}^{L}dz\,\delta V(%
{\bf r},z)\sin ^{2}(\frac{n\pi }{L}z)}{L/2}  \nonumber \\
&=&\frac{2e}{L}\int_{0}^{L}dz\,\delta \phi ({\bf q},z)\sin ^{2}(\frac{n\pi }{%
L}z),
\end{eqnarray}
where $\delta \phi ({\bf q},z)$ is of the order of $\Delta $. We shall
denote $\int_{0}^{L}dz\,\delta \phi ({\bf q},z)\sin ^{2}(\frac{n\pi }{L}z)$%
{\bf \ }by $\delta \tilde{\phi}({\bf q})$.

We now express the induced charge density in terms of $\delta \tilde{\phi}(%
{\bf q})$. This can be achieved in the conventional linear response theory
by 
\begin{equation}
\rho _{ind}({\bf q},\omega =0,z)=\int dz^{\prime }%
\mathop{\rm Re}%
\Pi (q,\omega =0,z,z^{\prime })\,e^{2}\delta \phi ({\bf q},z^{\prime }),
\end{equation}
where $\Pi (q,\omega =0,z,z^{\prime })\,$ is the polarization insertion\cite
{fetter}. If we focus on the n-th subband, the one-loop contribution to $\Pi
(q,\omega =0,z,z^{\prime })$ is 
\begin{equation}
\mathop{\rm Re}%
\Pi (q,\omega =0,z,z^{\prime })\,=\sin ^{2}(\frac{n\pi }{L}z)\sin ^{2}(\frac{%
n\pi }{L}z^{\prime })\frac{-16m}{\hbar ^{2}L^{2}}{\cal P}\int \frac{d^{2}k}{%
(2\pi )^{2}}\theta (1-k)\frac{1}{qk(\cos \theta +x)},  \label{polarization}
\end{equation}
where $q${\bf \ }and $k$ are measured in terms of $k_{F}$, $\theta (1-k)$ is
the step function, $x\equiv q/2k$, and ${\cal P}$ denotes the Cauchy
principle value. It is easy to show that 
\begin{equation}
{\cal P}\int_{0}^{2\pi }d\theta \,\frac{1}{\cos \theta +x}=\left\{ 
\begin{array}{c}
0\text{ \ \ \ \ \ \ \ if\ }|x|<1 \\ 
\frac{2\pi }{\sqrt{x^{2}-1}}\;\;\text{if }|x|\geq 1\;
\end{array}
\right. 
\end{equation}
Since the momentum transfer $q$\ is always less than $2k_{F}$, $x<1$\ for
the range of $k$ integration. We find that 
\begin{equation}
{\cal P}\int \frac{d^{2}k}{(2\pi )^{2}}\theta (1-k)\frac{1}{qk(\cos \theta
+x)}=\int_{0}^{q/2}\frac{dk}{2\pi }\frac{1}{q\sqrt{(q/2k)^{2}-1}}=\frac{1}{%
4\pi }\text{.}  \label{pi}
\end{equation}
As a result, we obtain 
\begin{equation}
\rho _{ind}({\bf q},\omega =0,z)=-\frac{4me^{2}}{L^{2}\pi \hbar ^{2}}\sin
^{2}(\frac{n\pi }{L}z)\delta \tilde{\phi}({\bf q}).  \label{induced}
\end{equation}
Substituting the above into Eq.(\ref{ming}) and performing Fourier
transformation on both ${\bf r}$ and $z$, we find 
\begin{eqnarray}
(q^{2}+k_{z}^{2})\,\delta \phi ({\bf q},k_{z}) &=&{\bf -}\frac{16me^{2}}{%
L^{2}\hbar ^{2}}\upsilon (k_{z})\delta \tilde{\phi}(q){\bf -}\frac{16\pi
en_{0}}{L^{2}}\Delta ({\bf q})\left[ u\left( k_{z}\right) +\upsilon (k_{z})%
\right]   \nonumber \\
&&+\frac{16me}{\pi L\hbar ^{2}}\frac{\partial E_{n}}{\partial L}%
\int\limits_{k\leq k_{F}}d^{2}{\bf k}\frac{\Delta ({\bf q})}{k^{2}-\left| 
{\bf k}-{\bf q}\right| ^{2}}\upsilon (k_{z}),
\end{eqnarray}
where $u\left( k_{z}\right) =\frac{n\pi }{2L}\int_{0}^{L}e^{ik_{z}z}z\sin
\left( \frac{2n\pi }{L}z\right) dz$ and $\upsilon
(k_{z})=\int_{0}^{L}e^{ik_{z}z}\sin ^{2}(\frac{n\pi }{L}z)dz$. The ${\bf k}$
integration has been done in Eq.(\ref{pi}). It is also easy to show that 
\begin{equation}
\delta \tilde{\phi}({\bf q)}=\int_{-\infty }^{\infty }\frac{dk_{z}}{2\pi }%
\delta \phi ({\bf q},k_{z})\upsilon ^{\ast }(k_{z})
\end{equation}
Substituting $\delta \phi ({\bf q},k_{z})$ in the above equation, we obtain 
\begin{eqnarray}
\delta \tilde{\phi}({\bf q)} &=&{\bf -}\int_{-\infty }^{\infty }\frac{dk_{z}%
}{2\pi }\frac{|\upsilon (k_{z})|^{2}}{q^{2}+k_{z}^{2}}\left[ \frac{16\pi
en_{0}}{L^{2}}\Delta ({\bf q})+\frac{16me^{2}}{L^{2}\hbar ^{2}}\delta \tilde{%
\phi}({\bf q)-}\frac{8me}{L\hbar ^{2}}\frac{\partial E_{n}}{\partial L}%
\Delta ({\bf q})\right]   \nonumber  \label{phi} \\
&&-\int_{-\infty }^{\infty }\frac{dk_{z}}{2\pi }\frac{v^{\ast }(k_{z})\cdot
u(k_{z})}{q^{2}+k_{z}^{2}}\frac{16\pi en_{0}}{L^{2}}\Delta ({\bf q})\text{.}
\label{phi}
\end{eqnarray}
The $k_{z}$ integration can be analytically solved as: 
\begin{eqnarray}
I(q) &\equiv &\int \frac{dk_{z}}{2\pi }\frac{|\upsilon (k_{z})|^{2}}{%
q^{2}+k_{z}^{2}}=\frac{L}{2}\left[ \frac{1}{q^{2}}+\frac{1}{2\left[
q^{2}+(2n\pi /L)^{2}\right] }\right] +\frac{(2n\pi /L)^{4}}{4q^{3}}\frac{%
e^{-qL}-1}{\left[ q^{2}+(2n\pi /L)^{2}\right] ^{2}} \\
J(q) &\equiv &\int \frac{dk_{z}}{2\pi }\frac{v^{\ast }u}{q^{2}+k_{z}^{2}} 
\nonumber \\
&=&\left( \frac{n\pi }{L}\right) ^{3}\left[ \frac{4}{qL}\frac{e^{-qL}-1}{%
\left[ q^{2}+(2n\pi /L)^{2}\right] ^{3}}+\frac{n\pi e^{-qL}}{q^{2}\left[
q^{2}+(2n\pi /L)^{2}\right] ^{2}}-\frac{n\pi }{q^{2}}\left( \frac{L}{2n\pi }%
\right) ^{4}\right] .
\end{eqnarray}
With a little rearrangement, Eq.(\ref{phi}) gives 
\begin{equation}
\delta \tilde{\phi}({\bf q)}=\frac{\frac{16\pi en_{0}}{L^{2}}\left[ I(q)+J(q)%
\right] -\frac{8me}{L\hbar ^{2}}\frac{\partial E_{n}}{\partial L}I(q)}{1+%
\frac{16me^{2}}{L^{2}\hbar ^{2}}I(q)}\left[ -\Delta ({\bf q})\right] 
\label{D}
\end{equation}
For a narrow quantum well satisfying $qL\leq 2k_{F}L\ll 1$ (this requires $%
L\ll 33$\r{A}\ for $n_{0}\approx 2\times 10^{15}$m$^{-2}$ in quantum wells), 
$I(q)\approx \frac{L^{2}}{8q}$ and $J(q)\approx -\frac{L^{2}}{16q}$ and Eq.(%
\ref{D}) reduces to the standard 2D screening form\cite{Ando1982}: 
\begin{equation}
\delta \tilde{\phi}({\bf q)}=\frac{\alpha }{q+q_{s}}\left[ -\Delta ({\bf q})%
\right]   \label{2D}
\end{equation}
(the second term in the denominator of Eq.(\ref{D}) does not exist in pure
two dimensions) where $q_{s}=2me^{2}/\hbar ^{2}\simeq 1/\left( 0.25\text{ \r{%
A}}\right) $ and $\alpha =\pi en_{0}$.

However, if $qL\approx 1$, one shall have to use the full expression of $I(q)
$ and $J(q)$. Since $\frac{16me^{2}}{L^{2}\hbar ^{2}}\simeq $10$^{30}$ m$%
^{-3}$ and $I(q)$ $\simeq {\rm 10}^{-27}{\rm m}^{3}$, the second term in the
denominator of Eq.(\ref{D}) dominates and $\delta \tilde{\phi}({\bf %
q)\approx }\frac{\pi n_{0}\hbar ^{2}}{2em}\Delta ({\bf q})$. The resulting
scattering matrix within a given subband is thus given by 
\begin{eqnarray}
|M({\bf q})|^{2} &=&\langle |\delta V({\bf q})|^{2}\rangle =\frac{1}{\aleph }%
\left( \frac{\partial E_{n}}{\partial L}-\frac{\frac{32\pi e^{2}n_{0}}{L^{3}}%
\left[ I(q)+J(q)\right] -\frac{16me^{2}}{L^{2}\hbar ^{2}}\frac{\partial E_{n}%
}{\partial L}I(q)}{1+\frac{16me^{2}}{L^{2}\hbar ^{2}}I(q)}\right) ^{2}S(q) 
\nonumber  \label{M2} \\
&\approx &\frac{1}{\aleph }\left( 2\frac{\partial E_{n}}{\partial L}-\frac{%
2n_{0}\pi \hbar ^{2}}{mL}\right) ^{2}S(q),  \label{M2}
\end{eqnarray}
where $\aleph ${\bf \ }is the normalization and $S(q)$ is the power spectrum
of $\Delta ({\bf q})$, given\cite{Penner} by $\langle |\Delta ({\bf q)}%
|^{2}\rangle $. Given the scattering matrix, we can calculate the relaxation
time via the relation 
\begin{equation}
\frac{1}{\tau ({\bf k})}=\frac{1}{2\pi \hbar }\int d^{2}k^{\prime }|M({\bf k}%
-{\bf k}^{\prime })|^{2}(1-\cos \Phi )\delta (E({\bf k})-E({\bf k}^{\prime
})),  \label{tau}
\end{equation}
where $\Phi $ denotes the angle between the initial and final wavevectors $%
{\bf k}$ and ${\bf k}^{\prime }$. The mobility of the electron can then be
solved by 
\begin{equation}
\mu =e\int dE\frac{\rho (E)\upsilon _{x}^{2}(E)\tau (E)}{4nk_{B}T\cosh
^{2}((E-E_{F})/2k_{B}T)}.  \label{mobility}
\end{equation}
We see that the interparticle interaction reduces the electron mobility
estimated by the energy level fluctuations by at least three quarters. Since 
$\frac{\partial E_{n}}{\partial L}<0$, the second term in Eq.(\ref{M2}) due
to wavefunction modulation increases the scattering matrix and further
reduces the mobility. The overall reduction of the mobility in comparison to
previous approach is about 1/4.

We note in passing that in general, in addition to the above Coulomb
interaction, the density modulation induced by the surface roughness also
affects any interactions that depend on the electron density. If in the
absence of surface roughness, the interaction is described by $\int d{\bf r}%
\int d{\bf r}^{\prime }{\rm \hat{n}}({\bf r}){\rm V}_0({\bf r,r}^{\prime })%
{\rm \hat{n}}({\bf r}^{\prime })$,\ then formally the effect of surface
roughness can be simply included by replacing ${\rm V}_0$ by 
\begin{equation}
{\rm V}({\bf r,r}^{\prime })\approx {\rm V}_0({\bf r,r}^{\prime })\left( 1-%
\frac{\Delta ({\bf r})+\Delta ({\bf r}^{\prime })}L\right) .
\end{equation}

\section{Discussions and conclusions}

In this section, we examine the validity of our approach by studying a
simplified rough interface: $z_{+}=L\cdot \theta (-x)+(L+\Delta )\cdot
\theta (x)$ and $z_{-}=0$, i.e., a step at $x=0$ (see Fig. 2). We shall
directly investigate the solution without using the effective Hamiltonian
Eq.(\ref{potent}) and show that they are consistent with each other. Let us
first fix the boundary condition by requiring the wavefunction to be
travelling waves when far from the step: i.e., $\Psi (x,z)\rightarrow
(e^{ikx}+{\cal R}e^{-ikx})\sin (\pi z/L)$ as $x\rightarrow -\infty $ and $%
\Psi (x,z)\rightarrow {\cal T}e^{ipx}\sin (\pi z/(L+\Delta ))$ as $%
x\rightarrow \infty $. When we approach $x=0$, higher subbands begin to
participate, and the most general form for the wavefunction is 
\begin{equation}
\Psi (x,z)=\left\{ 
\begin{array}{c}
(e^{ikx}+{\cal R}e^{-ikx})\sin \frac{\pi z}{L}+\sum\limits_{n=2}^{\infty
}b_{n}\,e^{k_{n}x}\sin \frac{n\pi z}{L}, \\ 
{\cal T}e^{ipx}\sin \frac{\pi z}{L+\Delta }+\sum\limits_{l=2}^{\infty
}a_{l}\,e^{-p_{l}x}\sin \frac{l\pi z}{L+\Delta },
\end{array}
\right. \quad \left. 
\begin{array}{c}
\text{for }x\geq 0 \\ 
\text{for }x<0
\end{array}
\right.   \label{psi}
\end{equation}
where $k_{n}$ and $p_{l}$ are positive and satisfy the energy conservation
condition: 
\begin{equation}
-k_{n}^{2}+\left( \frac{n\pi }{L}\right) ^{2}=-p_{l}^{2}+\left( \frac{l\pi }{%
L+\Delta }\right) ^{2}=k^{2}+\left( \frac{\pi }{L}\right) ^{2}=p^{2}+\left( 
\frac{\pi }{L+\Delta }\right) ^{2}.  \label{Eenergy}
\end{equation}
The fact that $\pi /L\ $is comparable to the Fermi momentum makes these
higher subbands (with $n,l$ greater than one) correspond to decaying modes
when $\Delta \ll L$. Also in this limit, since ${\cal R}$, $b_{n}$ and $a_{l}
$ vanish when $\Delta =0$, we expect them to be no greater than the order of 
$O(\Delta /L)$.

To find the coefficients in $\Psi (x,z)$, we apply the matching conditions
at $x=0$%
\begin{eqnarray}
\Psi (0^{+},z) &=&\left\{ 
\begin{array}{c}
\Psi (0^{-},z), \\ 
0,
\end{array}
\right. \quad \left. 
\begin{array}{c}
\text{for }0\leq z\leq L \\ 
\text{for }L\leq z\leq L+\Delta 
\end{array}
\right.   \label{BC1} \\
\partial _{x}\Psi (0^{+},z) &=&\partial _{x}\Psi (0^{-},z)\text{ for }0\leq
z\leq L\text{.}  \label{BC2}
\end{eqnarray}
By using the completeness relation of $\sin (\pi z/(L+\Delta ))$ and Eq.(\ref
{BC1}), we obtain 
\begin{equation}
a_{l}=(1+{\cal R}){\rm Y}(1,l)+\sum\limits_{n=2}^{\infty }b_{n}{\rm Y}(n,l),
\label{a}
\end{equation}
where $a_{1}\equiv {\cal T}$ and ${\rm Y}(n,l)$ is defined by 
\begin{eqnarray}
{\rm Y}(n,l) &\equiv &\frac{2}{L+\Delta }\int_{0}^{L}dz\sin \frac{n\pi z}{L}%
\sin \frac{l\pi z}{L+\Delta }  \nonumber \\
&=&\frac{2n(-1)^{n}}{L(L+\Delta )\pi }\frac{\sin \frac{\ell \pi L}{L+\Delta }%
}{\left( \frac{\ell }{L+\Delta }\right) ^{2}-\left( \frac{n}{L}\right) ^{2}}.
\label{Y}
\end{eqnarray}
This implies that ${\rm Y}(n,l)=O(\Delta /L)$ when $n\neq l$, while ${\rm Y}%
(n,l)=O(1)$ when $n=l$. This fact, when combined with Eq.(\ref{a}), yields 
\begin{equation}
{\cal T}=(1+{\cal R})Y(1,1)+O(\Delta ^{2}/L^{2}).  \label{1}
\end{equation}
Similarly, using the completeness relation of $\sin (\pi z/L)$ and Eq.(\ref
{BC2}), we obtain 
\begin{equation}
ik(1-{\cal R})=\frac{L+\Delta }{L}\left( ip{\cal T}{\rm Y}%
(1,1)-\sum\limits_{l=2}^{\infty }a_{l}\,p_{l}{\rm Y}(1,l)\right) ,
\end{equation}
which, to first order in $\Delta /L$, reduces to 
\begin{equation}
k(1-{\cal R})=\frac{L+\Delta }{L}p{\cal T}\,{\rm Y}(1,1)+O(\Delta
^{2}/L^{2}).  \label{2}
\end{equation}
By combining Eqs.(\ref{1}) and (\ref{2}), the coefficient ${\cal T}\ $can be
determined as $-\frac{\Delta }{2L}\left( 1+\frac{\pi ^{2}}{k^{2}L^{2}}%
\right) $. Note that the second term $-\frac{\Delta }{2L}$ reproduces
precisely the rescaling of wavefunction in Eq.(\ref{exp}), while the third
term $-\frac{\Delta \pi ^{2}}{2k^{2}L^{3}}$ is nothing but the same reduction
\cite{shankar} of transmitted amplitude in the 1-D Schrodinger equation with
a potential barrier of height, $\frac{\partial E}{\partial L}\Delta $ (the
energy level fluctuations in Eq.(\ref{potential})): 
\begin{equation}
{\cal T}=\frac{2}{1+\sqrt{1+(2\pi ^{2}\Delta /k^{2}L^{3})}}\simeq 1-\frac{%
\Delta \pi ^{2}}{2k^{2}L^{3}}.
\end{equation}
Although $\Delta $ is assumed to be positive in the above derivations, we
have checked that our conclusions remains valid when $\Delta $ is negative.

We have thus seen that the validity of our effective Hamiltonian approach
has been fully checked. We now briefly re-examine the effect due to the
change of the single-particle state in the ballistic regime. For a single
step as we considered in the above, according to the Landauer formula\cite
{Landauer}, the conductance due to the step is given by 
\begin{equation}
G=\frac{e^{2}}{\pi \hbar }\frac{T}{R},  \label{Lan}
\end{equation}
where $T$\ and $R$\ are transmission and reflection probability. A simple
analysis shows that the mobility is given by 
\begin{equation}
\mu =\frac{|e|}{\pi \hbar n_{0}}\frac{T}{R}\simeq \frac{|e|}{\pi \hbar n_{0}}%
\left[ \frac{L(1+\pi ^{2}/k_{F}^{2}L^{2})}{\Delta }-1\right] \simeq 4836%
\left[ \frac{L(1+\pi ^{2}/k_{F}^{2}L^{2})}{\Delta }-1\right] \frac{{\rm cm}%
^{2}}{{\rm V}\sec }
\end{equation}
When the system has many steps, we simply replace $\Delta $\ by $\sum \Delta
_{i}$. Since $\Delta $ is at the order of 3-4 \r{A}, we can estimate $\sum
\Delta _{i}$ by the same order. Therefore, $\mu $ is about $10^{4}-10^{5}%
{\rm cm}^{2}/{\rm V}\sec $ for $L=100$ \r{A}. This number when combined with
the contribution from Eq.(\ref{M2}) (\symbol{126}10$^{5}{\rm cm}^{2}/{\rm V}%
\sec $) predicts that the mobility is at the order of 10$^{4}{\rm cm}^{2}/%
{\rm V}\sec ,$\ in close to experimental result\cite{Yutani}.

In conclusion, we have derived an effective Hamiltonian for two dimensional
quantum wells with rough interfaces. It is checked to give consistent
results for an exactly solvable model. Two new terms are generated. The
first term is identified to the local energy level fluctuations, which was
introduced phenomenologically in the literature but the previous form is now
shown to be valid only when the Hamiltonian has one single length scale. The
effect of this term on the electron mobility has been discussed before. The
other term is a new finding, which is shown to modulate the wavefunction and
cause fluctuations in the charge density. We discuss its effects on the
reduction of the electron mobility both at the level of the single-particle
state and by including the many-particle interactions. An estimate of the
electron mobility is made and gives rises to correct order in comparison to
experimental data.

\section{Acknowledgments}

We gratefully acknowledge discussions with Professor Ming-Che Chang and
support from the National Science Council of the Republic of China under
Grant Nos. NSC 88-2112-M-007-009 (CYM), -010 (TMH),  and 89-2112-M-007-024
(CYM), -025 (TMH).\

\begin{center}
{\large {\bf Figure Captions}}:
\end{center}

\begin{description}
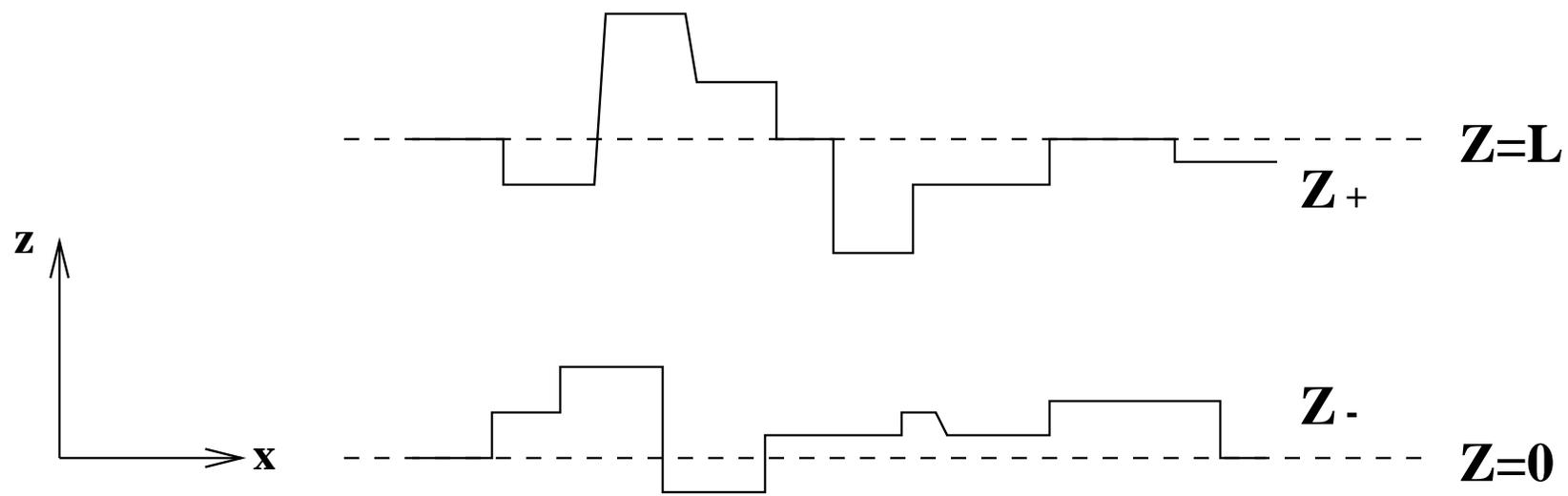
\item[Fig. 1]  A schematic plot of a 2D quantum well with rough interfaces

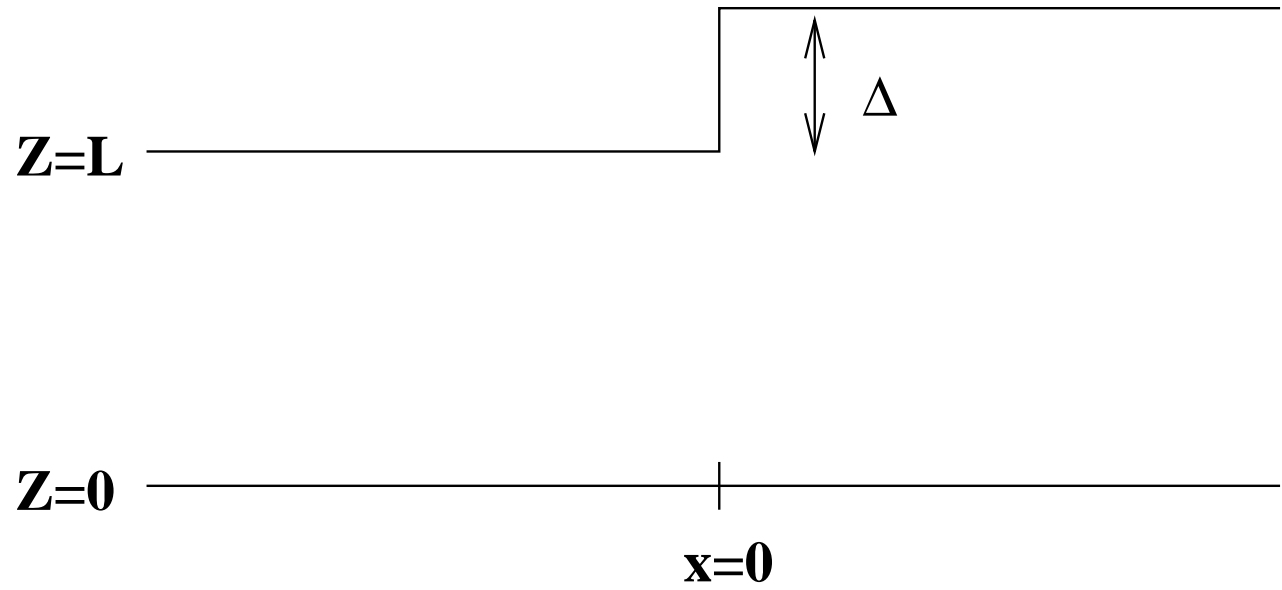
\item[Fig. 2]  A simple step at $x=0$ can be solved asymptotically.
\end{description}

\end{document}